\documentclass[aip,jcp]{revtex4-1}
\usepackage[pdftex]{graphicx}
\usepackage{amsmath}
\usepackage{amssymb}
\usepackage{color}

\begin{document}

\title{Ratcheted molecular-dynamics simulations identify efficiently the transition state of protein folding}
\date{\today}

\author{Guido Tiana}
\email{tiana@mi.infn.it}
\affiliation{Department of Physics, University of Milano, and INFN, via Celoria 16, 20133 Milan, Italy}
\author{Carlo Camilloni}
\email{cc536@cam.ac.uk}
\affiliation{Department of Chemistry, University of Cambridge, Lensfield Road, Cambridge, CB2 1EW, United Kingdom}

\begin{abstract}
The atomistic characterization of the transition state is a fundamental step to improve the understanding of the folding mechanism and the function of proteins. From a computational point of view, the identification of the conformations that build out the  transition state is particularly cumbersome, mainly because of the large computational cost of generating a   statistically--sound set of folding trajectories. Here we show that a biasing algorithm, based on the physics of the ratchet--and--pawl, can be used to identify efficiently the transition state. The basic idea is that the algorithmic ratchet exerts a force on the protein when it is climbing the free--energy barrier, while it is inactive when it is descending. The transition state can be identified as the point of the trajectory where the ratchet changes regime. Besides discussing this strategy in general terms, we test it within a protein model whose transition state can be studied independently by plain molecular dynamics simulations. Finally, we show its power in explicit--solvent simulations, obtaining and characterizing a set of transition--state conformations for ACBP and CI2.
\end{abstract}


\maketitle

\section{Introduction}

The transition state of biomolecular processes is particularly important because is the main determinant of the associated rate. Unfortunately, being the most unstable state of the process, its characterization is difficult. In the case of protein folding, a perturbative technique where one measures the effect of amino-acid mutations on folding/unfolding rates, has been successful in providing a structural characterization of this evanescent state \cite{Fersht:2002vx}. Although this procedure is experimentally rather demanding, we have now information about the structure of the transition state of tens of proteins.

With the improvement of the force fields \cite{Beauchamp:2012eq,LindorffLarsen:2012gl} that describes the interaction in proteins, it becomes more and more interesting the attempt to characterize the folding transition state without employing experimental information \cite{Calosci:2008dp,Geierhaas:2008gx,Paci:2002wo,Vendruscolo:2001em}. Within this context, the determination of the transition state implies two challenging problems, namely the generation of folding trajectories and the identification of the transition state along each of them. Concerning the former, the most straightforward way is simply to perform molecular--dynamics (MD) simulations solving the equation of motion of the system. In the case of proteins of realistic size and using realistic force fields in explicit solvent, generating a statistically--sound number of folding trajectories is not trivial even if one can use the fastest computers available \cite{LindorffLarsen:2011gl,Best:2012gn}. Smarter techniques, comprising transition path sampling \cite{Bol:2002}, milestoning \cite{Faradjian:2004cr} and dominant reaction pathways \cite{Faccioli:2006vz,Faccioli:2008wb}, exploit the fact that only a small subset of all possible trajectories is statistically relevant, but these methods are computationally efficient when the total number of atoms is not large (typically in implicit--solvent models). Even if one can generate efficiently folding trajectories, the problem of identifying the transition state is still hard. The transition state between two (meta)stable states is built out of the set of conformations for which the probability of falling down to each of them is 1/2. Consequently, the most direct way to identify the transition state is to start several MD simulations from each of the conformations selected from a folding trajectory, and to count the fraction of such trajectories which meet the native state before meeting the denatured state (or vice versa), until this fraction is exactly 1/2 \cite{Geissler:1999wy}. This procedure is very time consuming, but is the only safe way to identify the transition state \cite{Pande:1998tb}.

Some years ago, Marchi and Ballone introduced the idea of biasing MD simulations to generate efficiently trajectories between conformations of a system, using an algorithm based of the physics of the ratchet--and--pawl\cite{Mar:09}. It consists in defining a ratcheting coordinate ${\it y}$ and dumping the thermal fluctuations along the direction of ${\it y}$ opposite to the wished target. The algorithm was later used to enhance the thermal unfolding of proteins interacting with an implicit-solvent force field \cite{Paci:1999iu,Paci:2000ub}. Recently, ratcheted MD simulations were used to repeatedly simulate the folding of single--domain proteins in explicit solvent\cite{Camilloni:2011di}. Using a simplified protein model, a Beccara et al.\cite{ABeccara:2012fw} employed a Onsager--Machlup functional and showed that the ratcheted MD algorithm produces trajectories that are overall statistically relevant, thus validating the approach of ref. \cite{Camilloni:2011di}.

In what follows, we will investigate whether it is possible to use ratcheted MD simulations to obtain directly and efficiently a good approximation of the conformations which build out the transition state of protein folding. The basic idea is that while climbing the main free--energy barrier which separates the denatured from the folded state, the ratchet exerts work, while descending on the other side it is essentially off. The transition between the two regimes marks the transition state.

Although the whole goal of this work is to develop a method that can be used for realistic systems in explicit solvent, we first validate it using a model whose folding trajectories can be generated by plain MD and whose transition state can be obtained exactly using the committors method of ref. \cite{Geissler:1999wy}.

\section{The model and the simulations}

A model which is  suitable for developing a computational strategy and validating it against transition state obtained with the exact method is a modified all--atom G\=o model, where a non--specific interaction between hydrophobic atoms is added on the top of the native--structure. The G\=o model assures that folding can be simulated repeatedly also without the ratcheting algorithm, in order to be able to obtain reference trajectories, while the hydrophobic interaction makes the energy landscape more roughed, and thus more realistic. The G\=o implementation is that of ref. \cite{Whitford:2009ff}, in which pairs of atoms building native contacts interact with a Lennard--Jones potential whose minimum lies at $\epsilon_0=-0.62$ (in arbitrary units). The hydrophobic potential has also the Lennard--Jones form and acts between side chain carbons of ALA, VAL, LEU, ILE, PHE and TRP. The minimum of the potential lies at a distance of 0.35 nm, where the depth is $\epsilon_{hy}=-0.3$. This value of $\epsilon_{hy}$ has been chosen because it is the lowest which guarantees the folding of the proteins studied within an RMSD of 0.3 nm from the experimental native conformation. 

The simulations were carried out with a modified \cite{Cam:2008a,Bonomi:2009ul} version of Gromacs \cite{Hess:2008tf}, using the topologies generated with the SMOG web server \cite{Noel:2010jf}. The time step used is 0.002 ps (time units are merely nominal).

\begin{figure}
\includegraphics[width=\textwidth]{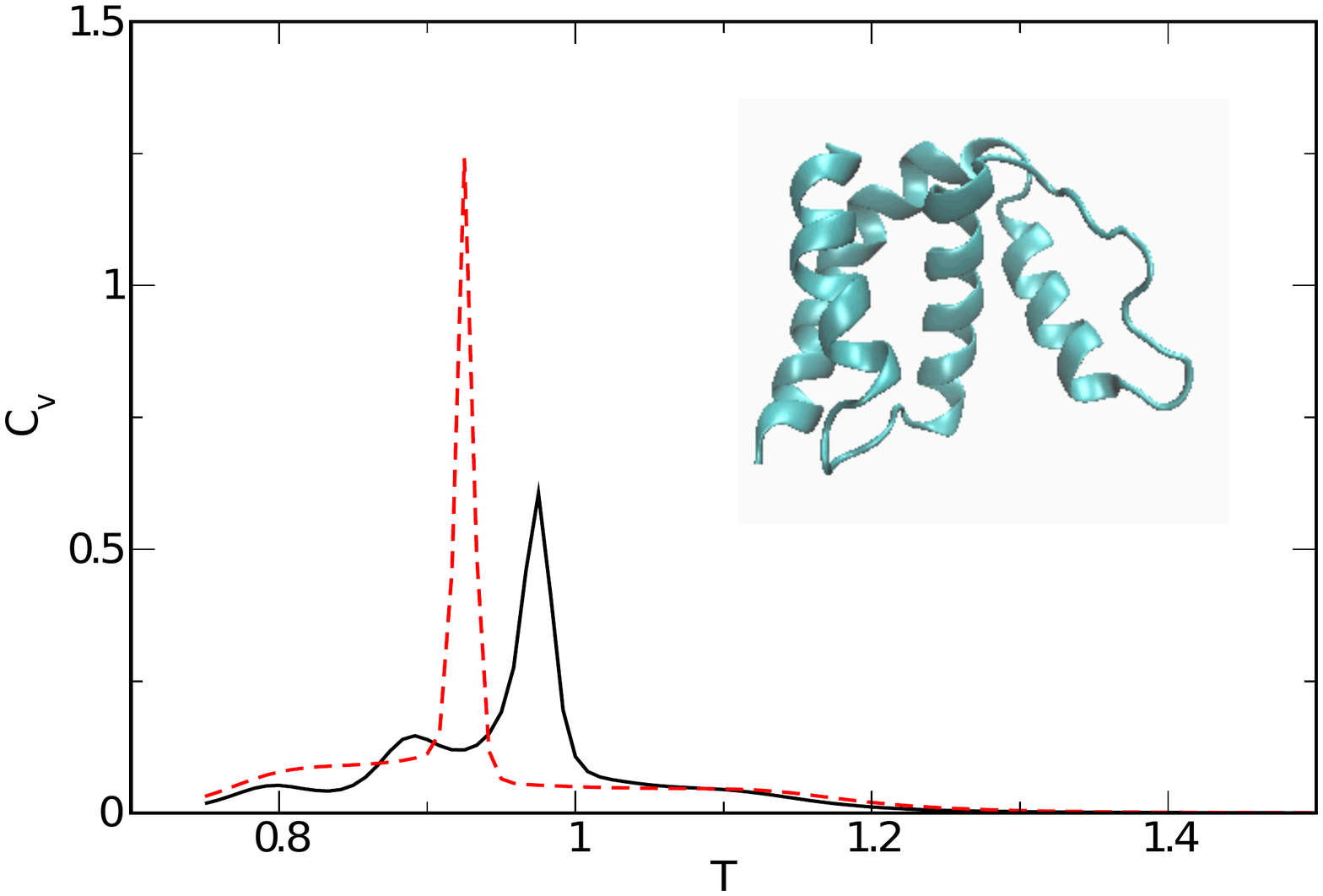}
\caption{The specific heat of ACBP (whose structure is displayed in the inset) as a function of temperature for the model interacting through the modified G\= o model (solid curve) and through a standard G\=o model (dashed curve). The temperature is expressed in energy units.}
\label{fig:cv}
\end{figure}

The specific heat, calculated with parallel--tempering simulations\cite {Hansmann:1997wb}, is displayed in Fig. \ref{fig:cv} and compared with that for a plain G\=o model. As expected, the two--states character of the denaturation transition diminished. However, the stability of the native state increased, suggesting that the hydrophobic interaction introduced in the model favors the native conformation, where hydrophobic packing is optimized, more than the denatured state. On the basis of this specific--heat plot, we used the trajectory obtained at $T=1$ to generate 10 uncorrelated unfolded conformations to be used as initial states of the folding simulations. The folding simulations were carried out at $T=0.91$, which is regarded as room temperature.

From each of the 10 unfolded conformations we carried out 10 simulations at $T=0.91$ for 6 ns each. The average folding time, defined as the time needed to reach a RMSD of 0.4 nm, is $\tau_f=1505$ ps.

Similar simulations were carried out using the ratcheting algorithm. The ratchet is implemented as in ref. \cite{Camilloni:2011di}, that is adding to the molecular potential a ratcheting term
\begin{equation}
V_{rat}(\rho(t)) =  \begin{cases} \frac{k}{2}\left(\rho(t)-\rho_m(t)\right)^2, &\rho(t)>\rho_m(t)\\
0, & \rho(t)\le\rho_m(t), \end{cases}
\end{equation}
where
\begin{equation}
\rho(t)=\left(y(t)-y_{target}\right)^2
\end{equation}
and
\begin{equation}
\rho_m(t)=\min_{0\le\tau\le t}\rho(\tau).
\end{equation}
The ratcheting coordinates $y(t)$ used in the present work are either the distance $d_{CM}$ of the contact map of a given protein conformation from the native contact map, or the RMSD (in both cases $y_{target}=0$). The distance $d_{CM}$, introduced by Bonomi et al.\cite{Bon:2007}, is defined as
\begin{equation}
\label{eq:dcm1}
d_{CM} = \|C-\tilde{C}\|=\left(\sum_{j>i+2}^{N}(C_{ij}-\tilde{C_{ij}})^2\right)^{1/2},
\end{equation}
were $C_{ij}$ is the i,j element of a NxN matrix defined as
\begin{equation}
\label{eq:dcm2}
C_{ij}(r_{ij}) =  \begin{cases} \frac{1-\left(\frac{r_{ij}}{r_0}\right)^p}{1-\left(\frac{r_{ij}}{r_0}\right)^q},&r_{ij} \leq r_{cut}\\
                                                     0, & r_{ij} > r_{cut}, \end{cases}
\end{equation}
$r_{ij}$ is the distance between atom $i$ and $j$ and $\tilde{C}$ is the defined on the native state.
The parameters used in these simulations are $p=6$, $q=10$, $r_0=0.75$ nm and $r_{cut}=1.23$ ${\rm nm}$.

Both in the case of plain--MD and ratcheted simulations, the sequence of events along the folding trajectories under each set of conditions were studied calculating the matrix $M_{ij}=\theta\left(t(i,k)-t(j,k)\right)$, where $t(i,k)$ is the  time at which the $i$th contact is stably formed in the $k$th simulation and $\theta$ is the Heaviside's step function. This matrix satisfies $M_{ij}+M_{ji}=1$ and each element $M_{ij}$  assumes the value $1$ if the formation of the $i$th contact precedes the formation of the $j$th, $0$ if it follows it, and $1/2$ if the two are uncorrelated. The average matrix
\begin{equation}
\overline{M_{ij}} = \frac{1}{n_s}\sum_{k=1}^{n_s} M_{ij},
\label{eq:m}
\end{equation}
where $n_s=100$ is the number of trajectories, is interpreted as the probability that the formation of the $i$th contact precedes the formation of the $j$th.  A  quantity related to $\overline{M_{ij}}$ is the probability $A_j=\sum_{i\neq j} \overline{M_{ij}}/(n_s-1)$ that the $j$th contact is formed after any other contact. 

The order of contact formation in two trajectories was compared using the distance
\begin{equation}
d(M,M')\equiv\frac{1}{n_s^2}\sum_{ij}[1-\delta(M_{ij},M'_{ij})],
\label{eq:d}
\end{equation}
between the associated matrices, where $\delta$ is the Kronecker symbol.

\section{Ratcheted trajectories}

A necessary condition for the ratcheting algorithm to identify the correct transition state of folding is to generate statistically--relevant trajectories. Failure of this condition  would lead to the identification of free--energy saddle points not corresponding to the main transition state of the folding process. Ratchet--generated trajectories are not expected to be associated -- as they are -- to a large statistical weight, because the corresponding folding time lies in the low--probability initial region of the folding--time distribution. However, as suggested in ref. \cite{Camilloni:2011di} and validated in this Section in the case of two model proteins, ratcheted MD simulations can provide the most probable sequence of contact formation if carried out in appropriate conditions. In this respect, ratcheted trajectories can be regarded as a coarse graining over time of the actual trajectories, in which the time--scale information is lost.

As a reference we generated 100 trajectories with plain MD simulations. The average folding time was 1505 ps and all trajectories reached the native conformations in the 10000 ps made available for each of them. The mean distance $\overline{d}$ between each pair of matrix $M_{ij}$ (cf. Eq. (\ref{eq:d})) is  0.37, indicating that the sequence of events along the different folding trajectories are rather homogeneous (cf. ref. \cite{Camilloni:2011di}). Briefly, this sequence implies first the formation of most contacts in the two terminal helices, than in the central helices and then the tertiary contacts.

\begin{figure}
\includegraphics[width=\textwidth]{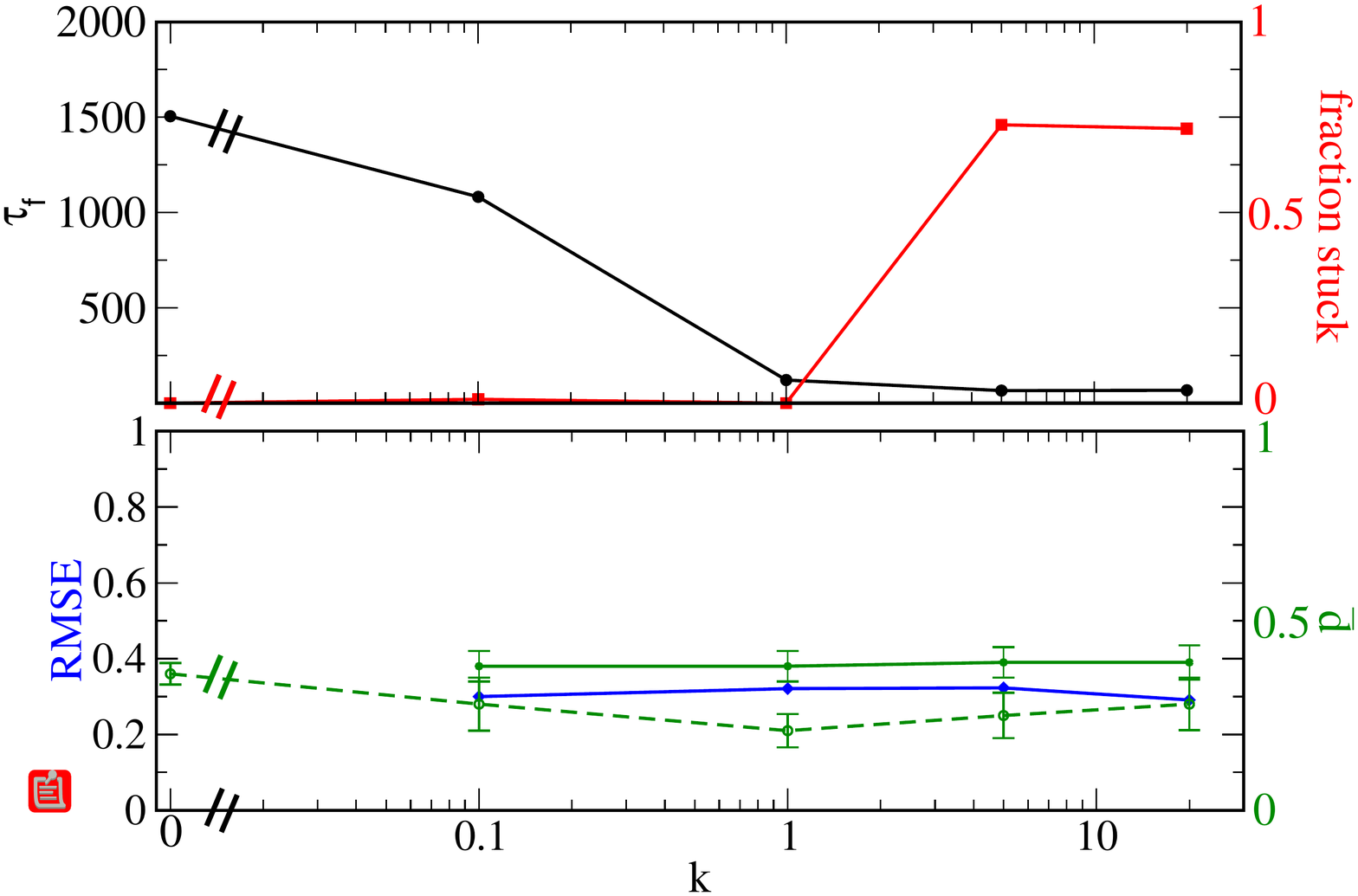}
\caption{Comparison of the folding simulations of ACBP ratcheted along $d_{CM}$ with those generated by plain MD. (upper panel) The average folding time (circles) and the fraction of stuck trajectories which are not able to reach the native state (squares), as a function of the ratcheting constant $k$. The latter is displayed in a logarithmic scale, except in the case of the points marked as $0$, which identify the simulation carried out without ratcheting. (lower panel) The root--mean--square error (RMSE) between the matrix $\overline{M_{ij}}$ calculated at $k$ and that calculated at $k=0$ (diamonds), the average distance $\overline{d}$ between the matrices $M_{ij}$ calculated at $k$ and  those calculated at $k=0$ (filled circles, the error bars indicate the standard deviation), and within the matrices $M_{ij}$ calculated at $k$ (empty circles). }
\label{fig:acbp_traj}
\end{figure}

Similar simulations were carried out starting from the same set of unfolded conformations, ratcheting the simulation along the distance $d_{CM}$ of the contact map to the native one with different values of the ratcheting constant $k$. Not all the trajectories folded to the native conformation, but some of them got stuck, reducing drastically the diffusivity of the different parts of the protein and, essentially, freezing to non--native conformations. These are excluded from the analysis that follows. The fraction of stuck trajectories, displayed in the upper panel of Fig. \ref{fig:acbp_traj}, increases with $k$. The same figure also displays the average folding time, which decreases as the effect of the ratchet is increased. The average folding time of ratcheted simulations has the only purpose of measuring the computational time needed to generate a folding trajectory, and has no physical meaning. The figure indicates that there is a range of values of $k$ around unity where simulations generate fast trajectories to the native conformations.

To assess the physical meaning of such trajectories, we  compared the order of native--contact formation to that of the unbiased simulations. The mean distance $\overline{d}$ between each pair of ratcheted trajectory  is around 0.3 for any value of $k$ and for the unbiased trajectories (see lower panel in Fig. \ref{fig:acbp_traj}), indicating that the sequence of events in the ratcheted simulations is as homogeneous as that of the MD trajectories. Also the mean distance between the matrices associated to ratcheted trajectories and those associated with plain--MD trajectories is 0.39 at all the values of $k$ considered. Comparing the average matrices $\overline{M_{ij}}$, one obtains that the root mean square error between the matrix $\overline{M_{ij}}$ generated ratcheting the simulations  and with plain MD is around 0.3 for all values of $k$ (cf. Fig. \ref{fig:acbp_traj}). 

Summing up, the difference between ratcheted and plain--MD trajectories is comparable with the (small) differences between pairs of plain--MD trajectories. Even when the ratchet is strong, although the fraction of folding trajectories drops drastically, the sequence of events in the few folding trajectories results correct.

\begin{figure}
\includegraphics[width=\textwidth]{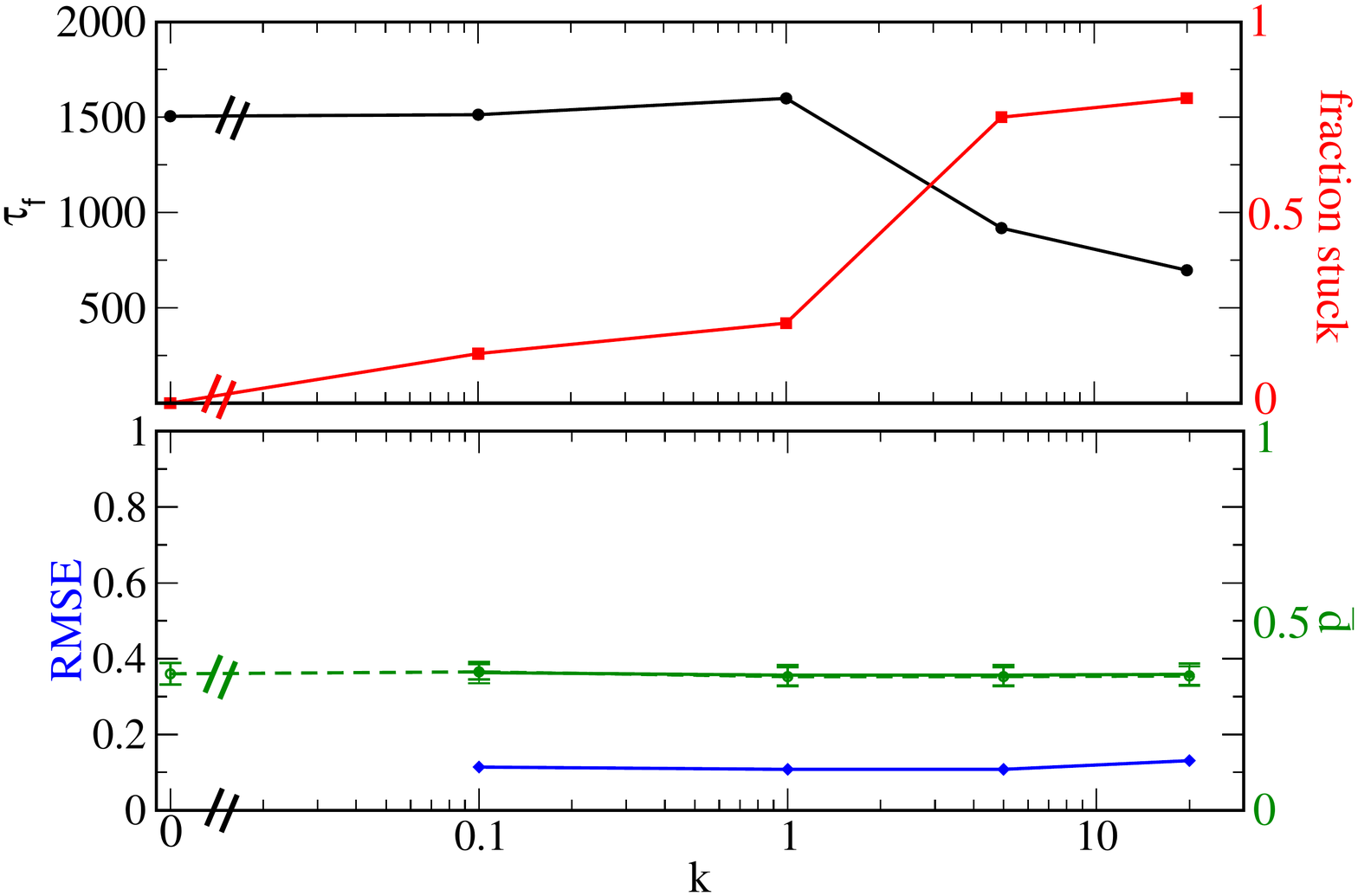}
\caption{Comparison of the folding simulations of ACBP ratcheted along the RMSD with those generated by plain MD. (upper panel) The average folding time (circles) and the fraction of stuck trajectories which are not able to reach the native state (squares), as a function of the ratcheting constant $k$. The latter is displayed in a logarithmic scale, except in the case of the points marked as $0$, which identify the simulation carried out without ratcheting. (lower panel) The root--mean--square error (RMSE) between the matrix $\overline{M_{ij}}$ calculated at $k$ and that calculated at $k=0$ (diamonds), the average distance $\overline{d}$ between the matrices $M_{ij}$ calculated at $k$ and  those calculated at $k=0$ (filled circles, the error bars indicate the standard deviation), and within the matrices $M_{ij}$ calculated at $k$ (empty circles). Here, the values of $k$ are given in energy units divided by nm. }
\label{fig:acbp_traj_rmsd}
\end{figure}

A similar analysis has been carried out ratcheting the simulation through a different coordinate, that is the RMSD with respect to the native conformation. Usually this is regarded as a bad reaction coordinate \cite{Best:2010cx}. In fact, attempts to fold small proteins in explicit solvent ratcheting along the RMSD coordinate at different values of the ratcheting constant have failed \cite{Camilloni:2011di}. The results of such simulations are displayed in Fig. \ref{fig:acbp_traj_rmsd}. Also in this case folding simulations display a sequence of events that is similar to the one generated by unbiased simulations. The main difference with the data obtained ratcheting along $d_{CM}$ is that in the present case there is not a range of values of $k$ at which ratchet is efficient. At small values of $k$ the folding time $\tau_f$ is essentially identical to that of the unbiased MD simulations. Only using values of $k$ larger than 10 one can observe a relevant decrease of $\tau_f$, but here the fraction of folding sequences has become negligible.

\begin{figure}
\includegraphics[width=\textwidth]{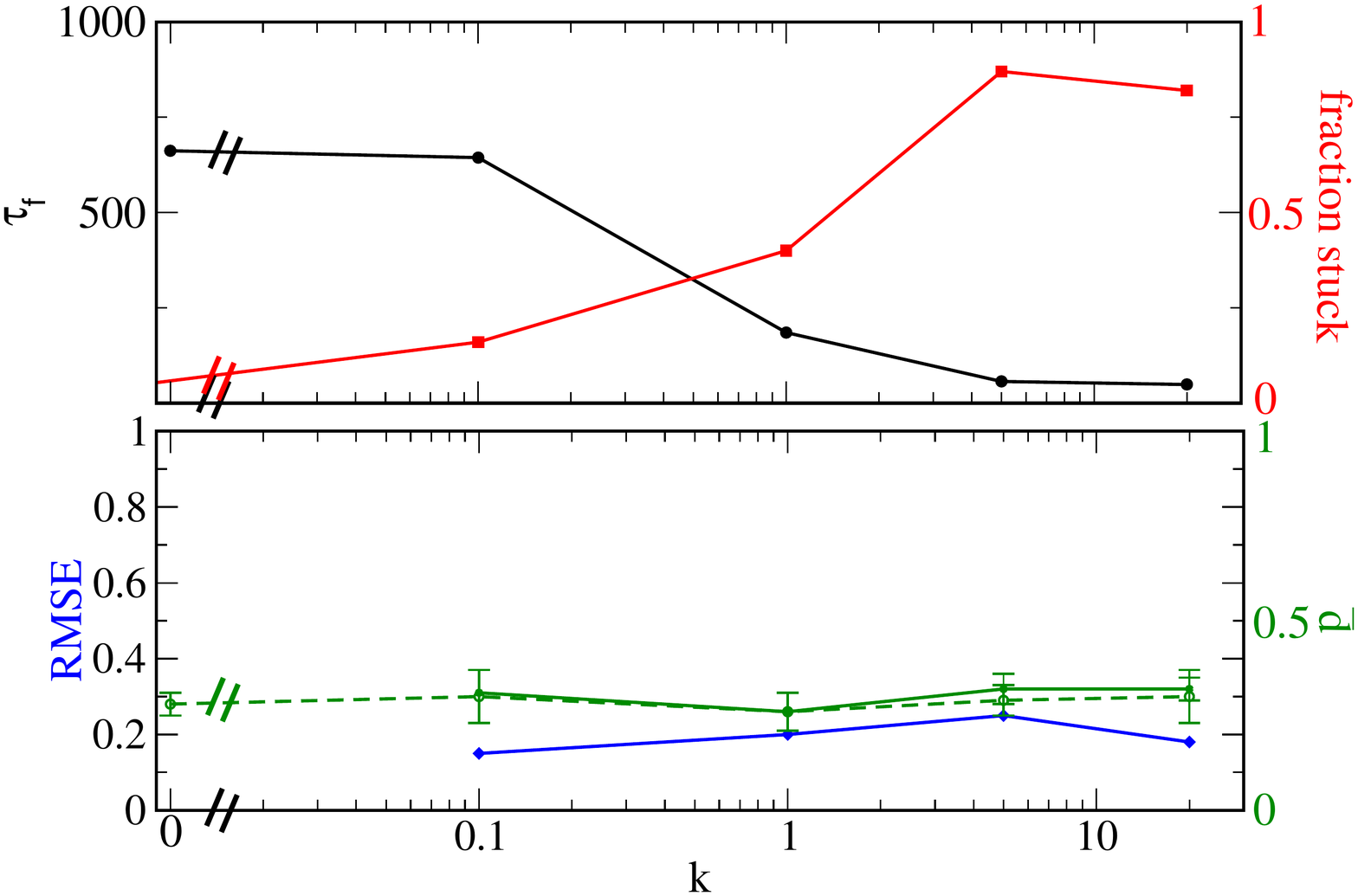}
\caption{Comparison of the folding simulations of CI2 ratcheted along $d_{CM}$ with those generated by plain MD. (upper panel) The average folding time (circles) and the fraction of stuck trajectories which are not able to reach the native state (squares), as a function of the ratcheting constant $k$. The latter is displayed in a logarithmic scale, except in the case of the points marked as $0$, which identify the simulation carried out without ratcheting. (lower panel) The root--mean--square error (RMSE) between the matrix $\overline{M_{ij}}$ calculated at $k$ and that calculated at $k=0$ (diamonds), the average distance $\overline{d}$ between the matrices $M_{ij}$ calculated at $k$ and  those calculated at $k=0$ (filled circles, the error bars indicate the standard deviation), and within the matrices $M_{ij}$ calculated at $k$ (empty circles).}
\label{fig:ci2_traj}
\end{figure}

ACBP is considered to fold according to a hierarchical diffusion--collision model \cite{Kragelund:1999il}, where first elements of secondary structure are formed, then diffuse around until they bind together to form native tertiary contacts. This pattern, which is also observed in the present simulations, could favor the applicability of the ratchet. To check the generality of the above results we have tested it with another case, namely CI2, which is considered the prototype of proteins which fold according to a nucleation model, without populating consistently secondary structures prior to the transition state. The results are displayed in Fig. \ref{fig:ci2_traj}. Also in this case there is a range of $k$ where the ratchet is efficient, that is both the folding time and the fraction of stuck trajectories are small. The efficiency is smaller than in the case of ACBP, probably because in this case the contact--map distance $d_{CM}$ is not a reaction coordinate as good as for a protein folding through a diffusion--collision scenario. Anyway, the sequence of events results to agree with that of a plain MD simulation, within the range of variability of the latter (which is somewhat small than that of ACBP).

\section{Identification of the transition state: the strategy}

The analysis of the time--dependence of the degrees of freedom associated with the ratchet can provide some information to localize the transition state of the system. The basic idea is that, as the system climbs the free energy barrier whose top is the transition state, the ratchet is very active and thus $V_{rat}$ is well above zero. When the system crosses the transition state and descends the free--energy barrier, the ratchet is essentially inactive and  $V_{rat}$ small. The point of the trajectory where  $V_{rat}$ drops is hypothesized to be the transitions state. 

Before verifying this hypothesis, we attempt to formalize the above idea, in a simple scenario where the molecular force can be approximated in an elementary form. Assuming that the dynamics of the degrees of freedom $\vec{x}$ of the system can be described by an over--damped dynamics
\begin{equation}
\frac{d\vec{x}}{dt}=\frac{1}{\gamma}\left[\vec{f}(\vec{x})-k\Delta\rho\cdot\vec{u}_\rho+\vec{\eta}\right],
\end{equation}
where $\gamma$ is the friction coefficient; $\eta$ is the thermal noise satisfying $<\vec{\eta}(t)\cdot\vec{\eta}(t')>=(2NT\gamma)\delta(t-t')$; $\vec{u}_\rho$ the versor that defines the direction of the ratcheting coordinate $\rho$; $\Delta\rho(t)\equiv\rho-\rho_m$ is the difference between the value of the ratcheting coordinate and its minimum; and Boltzman's constant is set to 1. Let's assume that $\rho$ is a good reaction coordinate, that is it moves according to the slowest time scale of the system \cite{Risken:1996vl}, and that the associated diffusion constant is approximately equal to that of the microscopic degrees of freedom. Then, the dynamics of $\rho$ can be written as
\begin{equation}
\frac{d\rho}{dt}=\frac{1}{\gamma}\left[  f_\rho-k\Delta\rho+\eta \right],
\end{equation}
where $f_\rho$ is the effective force which moves the one--dimensional degree of freedom $\rho$ (i.e., minus the gradient of the free energy).  By virtue of its definition, $\rho_m$ follows the dynamics
\begin{equation}
\frac{d\rho_m}{dt}=\frac{d\rho}{dt}\delta(\Delta\rho)\theta(-d\rho/dt),
\end{equation}
where $\theta$ is a step function that is 1 if its argument is positive and 0 otherwise. Consequently the quantity $\Delta\rho$ which measures the activity of the ratchet follows
\begin{equation}
\frac{d(\Delta\rho)}{dt}=
	\begin{cases}
	  \frac{1}{\gamma}\left[f_\rho-k\Delta\rho+\eta\right] & \text{if } \Delta\rho>0 \text{ or } d\rho/dt>0 \\
	 0 & \text{if } \Delta\rho=0 \text{ and } d\rho/dt<0.
	\end{cases}
\label{eq:dr}
\end{equation}
If the molecular force $\vec{f}$ pushes the system downhill towards its target state and it is overwhelming with respect to the typical diffusive force (i.e., $f_\rho\ll -(2T\gamma/\Delta t)^{1/2}$), then $\Delta\rho$ is approximately zero along the associated part of trajectory. 

A more common scenario is that where the system is running downhill, the diffusive term is not negligible, but the molecular force is overwhelming with respect to the ratchet (i.e., $f_\rho\ll -k\Delta\rho$), so that we can neglect the term proportional to $k$ in Eq. (\ref{eq:dr}).  To make things simple, let's focus on a fraction of the trajectory that is short enough that the force $f_\rho$ can be approximated as constant. In this case $\Delta\rho$ experiences a diffusion biased by a constant force, and by a trap at 0. In fact, when $\Delta\rho=0$ the system can move away only if $d\rho/dt>0$ (cf. the condition controlling the first line of Eq. (\ref{eq:dr})); thus the exit rate $w_{exit}$ from the trap is proportional to the probability that $\eta>|f_\rho|\,(\Delta t/2T\gamma)^{1/2}$. This case is analogous to that of a massive particle diffusing on a slope with a trap at the bottom. Since the stochastic noise $\eta$ is normally distributed, $w_{exit}$ is proportional to $\text{erfc}(|f_\rho|\,(\Delta t/2T\gamma)^{1/2})$. We can assign to the trap an effective energy $U_{trap}$, so that $w_{exit}$ is equal to Kramers escape rate, that is
\begin{equation}
\exp\left[\frac{U_{trap}}{T}\right]=\frac{1}{2}\text{erfc}\left[|f_\rho|\left(\frac{\Delta t}{2T\gamma}\right)^{1/2}\right].
\end{equation}
In the neighborhood of 0, $\Delta\rho$ will soon populate a distribution given by
\begin{equation}
p(\Delta \rho)= \begin{cases}
			\frac{2}{Z}\text{erfc}\left[|f_\rho|\left(\frac{\Delta t}{2T\gamma}\right)^{1/2} \right]^{-1} &\text{if }0<\Delta\rho<\epsilon \\
			\frac{1}{Z}\exp\left[-\frac{|f_\rho| \Delta\rho}{T}\right] &\text{if } \Delta\rho>\epsilon,
		   \end{cases}
\label{eq:py1}
\end{equation}
where $\epsilon$ is the (small) length which defines the trap and
\begin{equation}
Z=2\cdot\text{erfc}\left[|f_\rho|\left(\frac{\Delta t}{2T\gamma}\right)^{1/2} \right]^{-1} +\frac{T}{|f_\rho|\epsilon}.
\end{equation}
The average value of $\Delta\rho$ expected in this regime is then
\begin{equation}
<\Delta\rho>=\frac{T^2}{\epsilon f_\rho^2\,\text{erfc}[|f_\rho|(\Delta t)/2T\gamma)^{1/2}]^{-1}+|f_\rho|T} \xrightarrow{\epsilon\to 0}  \frac{T}{|f_\rho|}.
\end{equation}

On the other  hand, if the system is climbing the free--energy barrier (i.e. $f_\rho\gg(2T\gamma/\Delta t)^{1/2}$), the conditions $\Delta\rho=0$ and $d\rho/dt<0$ in Eq. (\ref{eq:dr}) are never satisfied simultaneously. Consequently,
\begin{equation}
p(\Delta\rho)=\frac{1}{Z}\exp\left[-\frac{\frac{k}{2}\Delta\rho^2-f_\rho \Delta\rho}{T}\right] 
\label{eq:py2}
\end{equation}
where
\begin{equation}
Z=\left(\frac{\pi T}{2k}\right)^{1/2}  \exp\left[\frac{f_\rho^2}{2kT}\right]  \left( 1+\text{erf}\left[\frac{f_\rho}{(2kT)^{1/2}} \right]\right),
\end{equation}
giving the average
\begin{equation}
<\Delta\rho>=2\frac{(2/\pi)^{1/2}Tk  \exp\left[ -\frac{f_\rho^2}{2kT}  \right] +f_\rho(Tk)^{1/2}   \left(1+\text{erf}\left[ \frac{f_\rho}{(2kT)^{1/2}} \right]\right)   }{(k^3T)^{1/2}\,\left(1+\text{erf}\left[ \frac{f_\rho}{(2kT)^{1/2}} \right]\right)  }    \xrightarrow{f_\rho^2\gg 2kT}  \frac{2f_\rho}{k} .
\end{equation}

The transition state is the intermediate scenario where $f_\rho$ vanishes. Assuming $|f_\rho|\ll 2T\gamma/\Delta t)^{1/2}$, one can neglect the molecular force in Eq. (\ref{eq:dr}), obtaining
\begin{equation}
p(\Delta \rho)= \begin{cases}
			2/Z &\text{if }0<\Delta\rho<\epsilon \\
			\frac{1}{Z}\exp\left[-\frac{k \Delta\rho^2}{2T}\right] &\text{if } \Delta\rho>\epsilon,
		   \end{cases}
\label{eq:py3}
\end{equation}
where $Z=2+(\pi k/2T)^{1/2}/\epsilon$. This is an ideal distribution, because it is unlikely that the system spends enough time at the transition state to populate it. However, it can be useful to obtain the average $\Delta\rho$ which separates the rising from the descending regime. In fact, we get
\begin{equation}
<\Delta\rho>=\frac{2T}{4k\epsilon+(2\pi kT)^{1/2}}\xrightarrow{\epsilon\to 0}\left(\frac{2 T}{\pi k} \right)^{1/2}.
\label{eq:drho}
\end{equation}

\begin{figure}
\includegraphics[width=\textwidth]{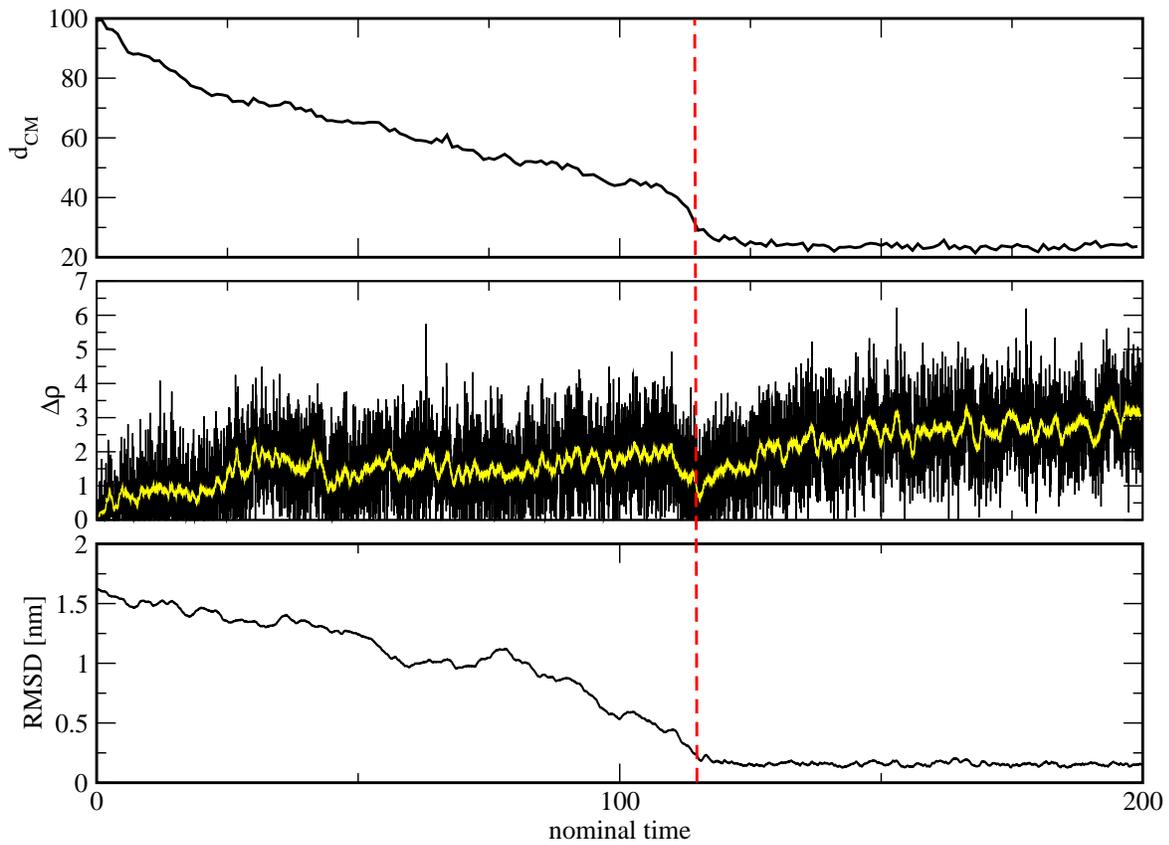}
\caption{The typical behavior of the contact--map distance to the native conformation $d_{CM}$, the ratchet displacement $\Delta\rho$ and the RMSD to the native conformation in a folding trajectory ratcheted with $k=1$. The light curve in the middle panel is a 0.8--ps running average of the underlying curve. The vertical dashed bar marks the TS identified in they trajectory.}
\label{fig:ts_example}
\end{figure}

\begin{figure}
\includegraphics[width=\textwidth]{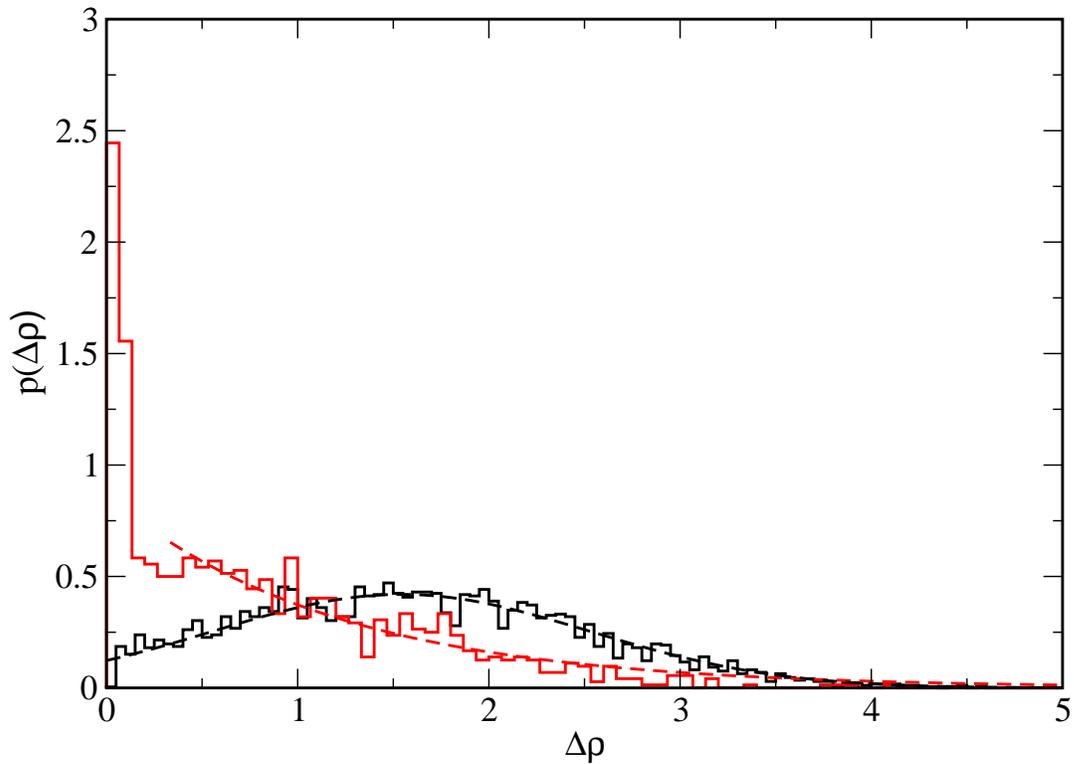}
\caption{The histogram of $\Delta\rho$ obtained from the parts of the trajectory of Fig. \protect\ref{fig:ts_example} which ranges between nominal times 30 and 100 ps, presumably corresponding to the climbing of the folding free--energy barrier (black solid curve) and between 110 and 125 ps, presumably corresponding to the descent to the native state (red solid curve). The fit obtained by Eq. \protect\ref{eq:py2} (dashed black curve) and   \protect\ref{eq:py1} (dashed red curve) are also indicated. }
\label{fig:fit}
\end{figure}

The behavior of $\Delta\rho$ in a typical ratcheted simulation is displayed in the middle panel of Fig. \ref{fig:ts_example}. Although it is difficult to distinguish {\it a priori} where the system is climbing and where it is descending the folding free--energy barrier, it is reasonable to argue that in part of the trajectory in the range $30<t<125$ the system is climbing, while in the range $110<t<125$ it is descending. The distribution $p(\Delta\rho)$ associated with these two parts of the trajectory are displayed in Fig. \ref{fig:fit} with solid black and red curves, respectively. The black curve is fitted by Eq. (\ref{eq:py2}), the correlation coefficient being $0.958$. The red curve displays a sharp peak at low values of $\Delta\rho$ as predicted by the first line of Eq. (\ref{eq:py1}), allowing to obtain $\epsilon=0.3$, while the remaining part is fitted by the second line of f Eq. (\ref{eq:py1}), with a correlation coefficient of $0.965$. This means that, although the molecular force $f_\rho$ certainly depends on the specific point of the trajectory, the system crosses the free-energy barrier experiencing an effective force of $f_\rho=0.75$ and descend it pushed by an effective force $f_\rho=-1.45$.

The value of $<\Delta\rho>$ obtained in Eq. (\ref{eq:drho}) can be used to estimate the order of magnitude of the threshold to distinguish the regime where the system is climbing the free-energy barrier from that in which it is descending, that is the transition state. For example, in the simulation we performed with $T=0.91$ and $k=1$, we obtain  $<\Delta\rho>=0.76$ (cf. Fig. \ref{fig:ts_example}).

\section{Identification of the transition state: results}

\begin{figure}
\includegraphics[width=\textwidth]{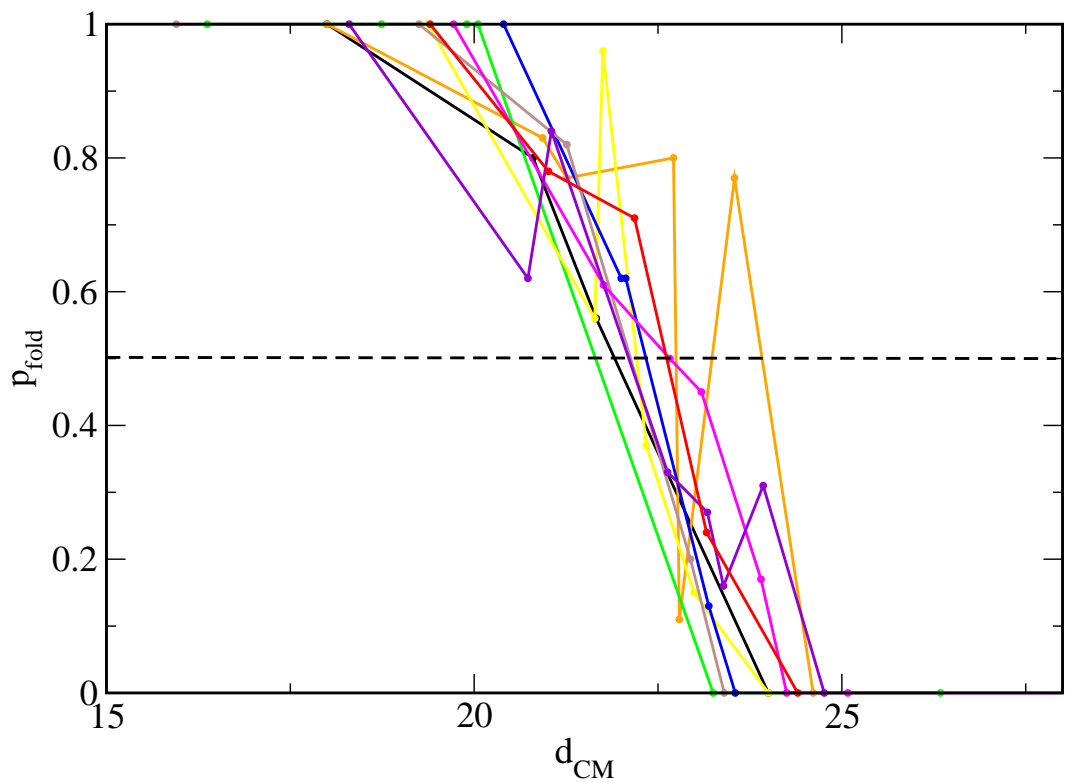}
\caption{The folding probability $p_{fold}$ calculated over a set of conformations extracted from 10 folding simulations. The conformations displaying $p_{fold}=0.5$ build out, by definition, the folding transition state. }
\label{fig:commit}
\end{figure}

\begin{figure}
\includegraphics[width=\textwidth]{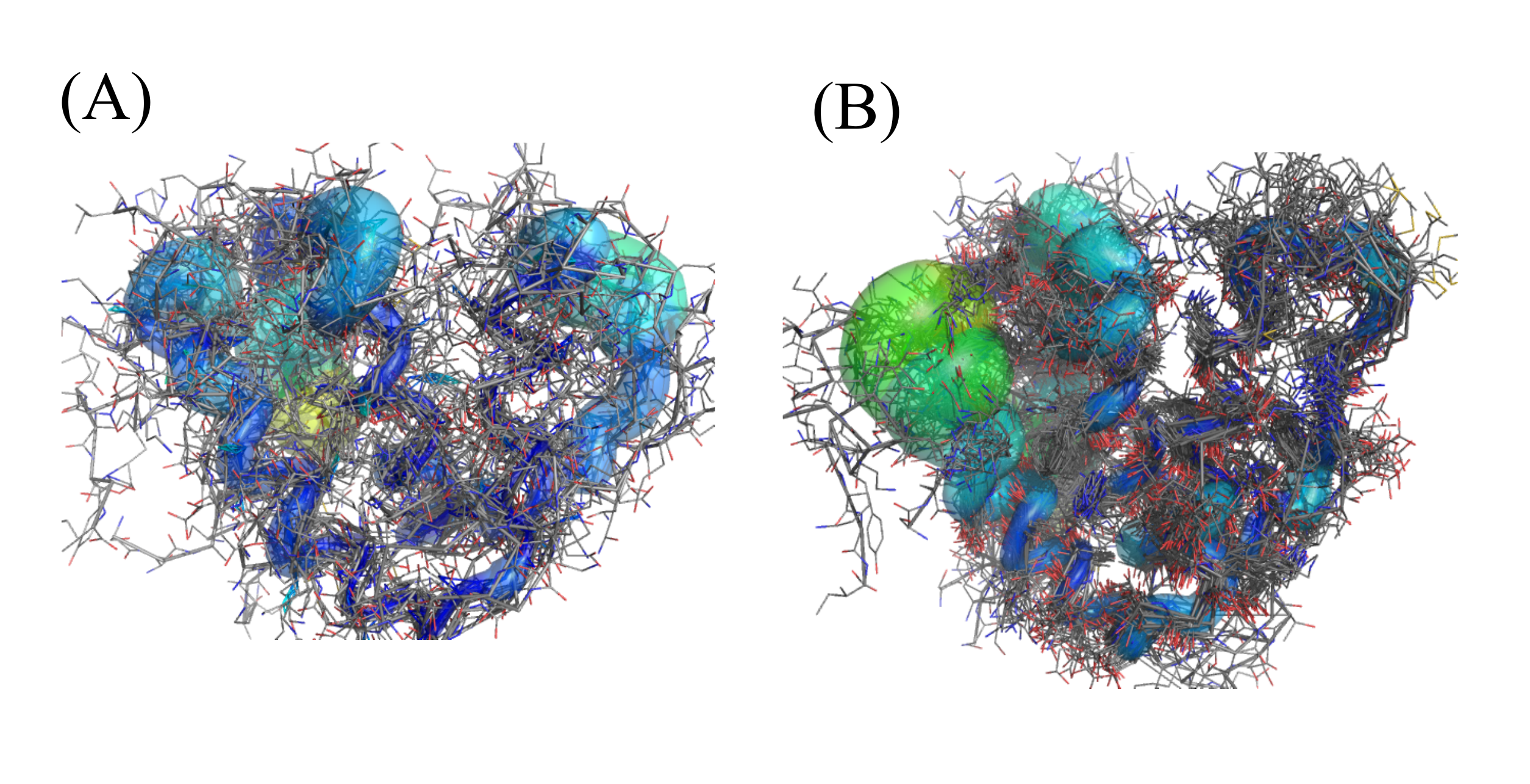}
\caption{Comparison between the G\=o--model transition--state conformations for ACBP obtained by plain--MD simulations through the commitment analysis (A) and those obtained by ratcheted simulations as explained in the text (B). The width and the color of the average conformations denote the RMS fluctuations.}
\label{fig:ts_go}
\end{figure}

Before applying the strategy discussed above, the actual TS was identified through a commitment analysis  \cite{Geissler:1999wy} on 10 plain--MD folding trajectories of ACBP. From each of them we extracted a variable number (from 5 to 10) of conformations chosen in the region where the value of $d_{CM}$ displays a rapid decrease to low values. From each of them we started 100 plain-MD simulations, calculating the probability $p_{fold}$ that the simulation reaches the native basin (operatively defined from $d_{CM}<19$) before reaching the denatured basin  (operatively defined from $d_{CM}>25$). The conformations displaying $0.4<p_{fold}<0.6$ are defined as TS conformations. The behavior of $p_{fold}$ with respect to the value of $d_{CM}$ of the associated conformation is displayed in Fig. \ref{fig:commit}. The associated conformations are displayed in Fig. \ref{fig:ts_go}(A). They are remarkably native--like, displaying an average RMSD to the native conformation of $0.68\pm 0.17$ nm, and fairly homogeneous, their mutual average RMSD being $0.85\pm 0.17$ nm.
 
For each trajectory generated with the ratcheting algorithm,  we have looked for the TS in the region where the RMSD to the native conformation was in the range between 0.2 nm and 1 nm. The putative TS is the conformation such that the average value of $\Delta\rho$ in the preceding 8 ps is larger than that predicted by Eq. (\ref{eq:drho}) and in the following 0.8 ps is smaller. In this way, we could identify a conformation in 64\% of trajectories at $k=0.1$, in the 86\% of the trajectories at $k=1$ and in the 49\% of trajectories at $k=20$. In no cases more than one conformation is identified.

\begin{figure}
\includegraphics[width=\textwidth]{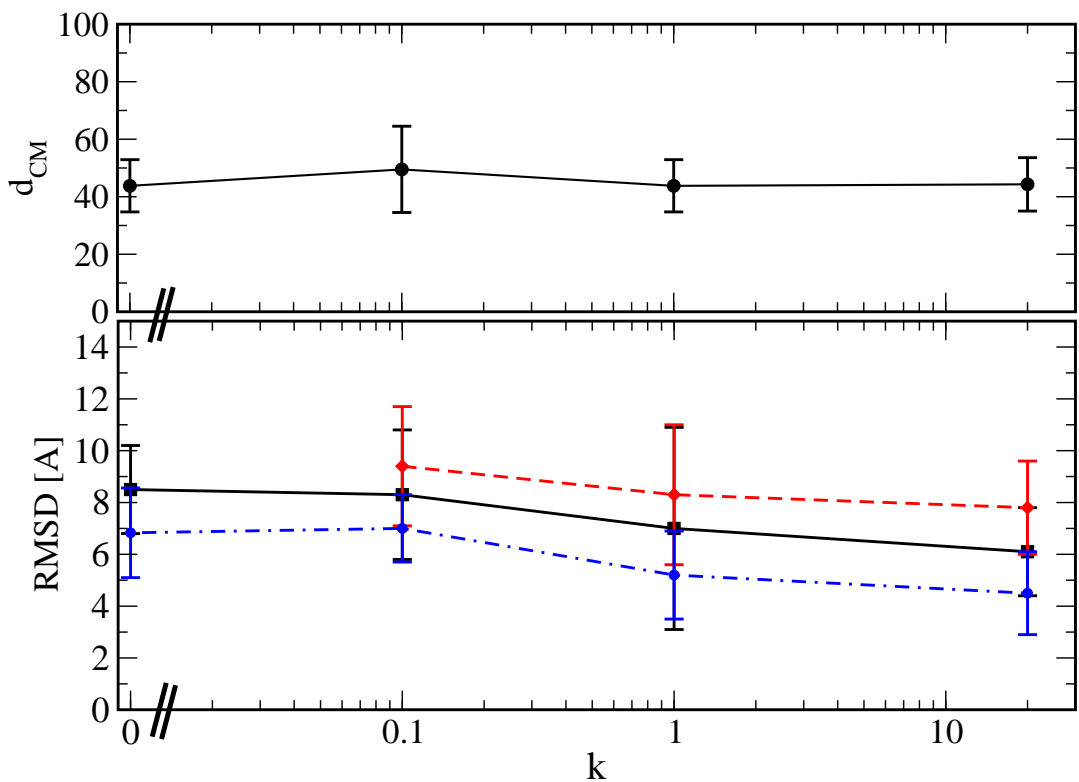}
\caption{Structural properties of transition--state conformations obtained at various values of $k$ and obtained from plain MD simulations ($k=0$). (upper panel) The average distance between the contact map of TS conformations and that of the native state. (lower panel) The average RMSD between pairs of TS conformations at each value of $k$ (black squares), the average RMSD between TS conformations obtained at different values of $k$ and those obtained by plain MD simulations (red diamonds) and average RMSD to the native conformation (blue circles). The error bars indicate the standard deviation.}
\label{fig:ts_k}
\end{figure}

The structural properties of the conformations identified by the above criteria are summarized in Fig. \ref{fig:ts_k}. The average contact--map distance is comparable to that of the actual TS at all values of $k$. The structural homogeneity of the TS conformations is slightly decreasing with the increasing of $k$, the mutual average RMSD going from $0.85$ nm at $k=0$ to $0.61$ nm at $k=20$. The average similarity of the TS conformations obtained from ratcheted simulations to the actual TS conformations is within the error bars $\sigma$ associated with the intrinsic variability of the TS conformations (the difference between the two averages being $\approx 0.2\sigma$; black error bars in the figure). Also the RMSD to the native conformation displays a slight decrease from $0.68$ nm at $k=0$ to $0.46$ nm at $k=20$. Summing up, at all values of $k$ analyzed, ratcheted MD simulations can identify TS conformations which are structurally similar to the actual TS conformations, becoming slightly more native--like at increasing $k$.

A representation of the protein in the TS obtained at $k=1$ is displayed in Fig. \ref{fig:ts_go}B. The main differences between the actual TS and that obtained by ratcheted conformation at $k=1$ involves the terminals of the protein. The actual TS displays large fluctuation in the C-terminal part of the chain and, to a smaller extent, in the N--terminal and in the loop region. The ratcheted TS overestimates the fluctuations in the C-terminal region, while it slightly underestimates those involving the loop. Anyway, the two sets are remarkably similar.

\section{An explicit--solvent case: the transition state of ACBP and CI2 simulated with the Amber force field}

The very goal of the strategy discussed above is not the identification of the transition state with simplified protein descriptions, but in realistic explicit--solvent models. In order to test the algorithm we analyzed the folding and unfolding trajectories generated using the ratcheting algorithm in ref. \cite{Camilloni:2011di}. Using the Amber03 force field \cite{Dua:03}, we simulated 10 folding and 10 unfolding trajectories of ACBP and CI2 in a dodecahedron box of 261 ${\rm nm}^3$  solvated with $\sim10^4$ TIP3P water molecules, ratcheting along $d_{CM}$ with a ratcheting constant $k=1 kJ/mol$ for 50 ns at $T=300K$. All trajectories folded within $0.25$ nm from the native conformation.

\begin{figure}
\includegraphics[width=\textwidth]{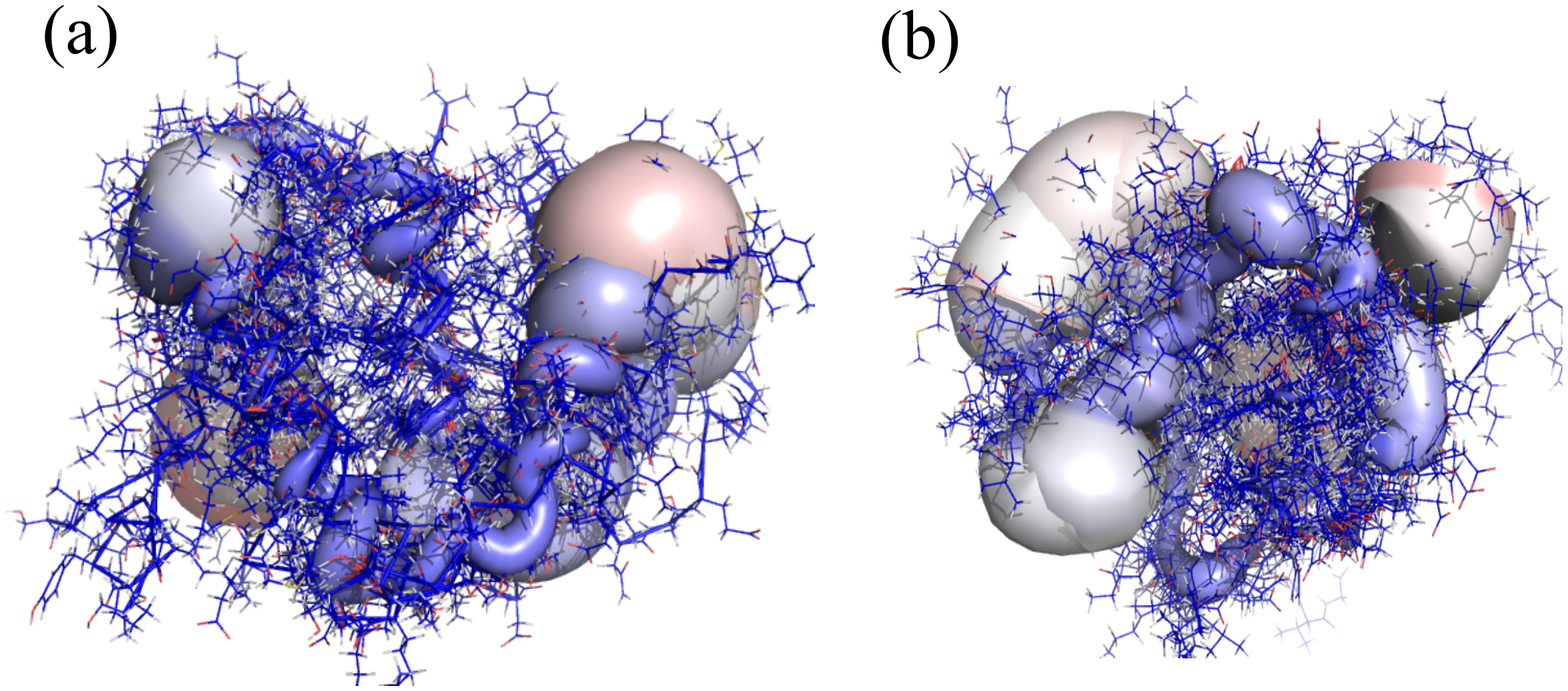}
\caption{The conformations corresponding to the transition state of ACBP (a) and CI2 (b) from the explicit--solvent ratcheted simulations. The thickness of the  surface indicates the standard deviation associated to the average structure.}
\label{fig:ts_solv}
\end{figure}

The transition state is identified with the same strategy used in the G\=o--model simulations, requiring that the average of $\Delta\rho$ in the preceding 8 ps is larger than 1 and in the following 8 ps is smaller than 2 (this is a somewhat looser condition than for the G\=o model, but guarantees the identification of a unique TS for each trajectory), while the RMSD to the native state should range between 0.3 and 1 nm. The conformation thus obtained are displayed in Fig. \ref{fig:ts_solv}. They are less homogeneous than those obtained by means of the G\=o model, the average mutual RMSD being $0.82\pm 0.19$ nm in the case of ACBP and $0.76\pm 0.14$ nm in the case of CI2.  Their RMSD to the native state is $0.68\pm 0.19$ nm in the case of ACBP and $0.70\pm 0.17$ nm in the case of CI2.

In order to validate the TS without carrying out a commitment analysis which is extremely time--consuming in explicit solvent \cite{Shaw:2010ge}, we have compared the TS conformations obtained from the folding trajectories to the TS conformations obtained by unfolding trajectories under the same conditions. According to the principle of detailed--balance, under the same conditions the two TS must be identical \cite{Finkelstein:1997tp}. 
The TS conformations obtained in this case are slightly more native--like, displaying a RMSD to the native conformation of $0.43\pm 0.13$ nm for ACBP and $0.62\pm 0.07$ nm for CI2. In order to compare the set of TS conformations obtained from folding and from unfolding trajectories, we have calculated the average pairwise RMSD of conformations across the two sets, which is $0.81\pm 0.10$ nm for ACBP and $0.76\pm 0.13$ nm for CI2. 

The average similarity between the folding and the unfolding TS is compatible, within the error bars, to the intrinsic heterogeneity of each set (their difference is $0.05\sigma$ in the case of ACBP and $0$ in the case of CI2),  and  so guarantees that the two TS can be regarded as approximatively identical. 

\section{Conclusions}
The complexity of  the characterisation of biomolecular processes is driving a continuos improvement of the experimental and the computational techniques \cite{Bartlett:2009ug}. In particular, in the field of computer simulations, in the last few years we have assisted in a leap in the accessible time scale of plain MD simulations.  Nonetheless even these major improvements are not able to address the complexity of folding problem for realistic proteins \cite{Shaw:2010ge}. This points to the necessity of carrying on with the development of both simplified model and advanced sampling methods. The present work further validates the use of the ratcheting algorithm in the study of protein folding and extends its use to the approximate identification, at an atomic level, of the transition state ensemble of a protein in explicit solvent.

\section{Acknowledgments}
CC was supported by a Marie Curie Intra-European fellowship. We acknowledge the use of computing facilities provided by CamGrid.


%

\newpage

\end{document}